\documentclass[floatfix,aps,pre,showpacs]{revtex4}
\usepackage{graphicx}
\usepackage{amssymb,amsfonts,amsmath}
\usepackage{bbm}
\begin{document}

\title{Classification and sparse-signature extraction from gene-expression data}

\author{Andrea Pagnani} \affiliation{Institute for Scientific
  Interchange, Viale Settimio Severo 65, Villa Gualino, I-10133
  Torino, Italy}

\author{Francesca Tria} \affiliation{Institute for Scientific
  Interchange, Viale Settimio Severo 65, Villa Gualino, I-10133
  Torino, Italy}

\author{Martin Weigt} \affiliation{Institute for Scientific
  Interchange, Viale Settimio Severo 65, Villa Gualino, I-10133
  Torino, Italy}

\date{\today}


\begin{abstract}
In this work we suggest a statistical mechanics approach to the
classification of high-dimensional data according to a binary
label. We propose an algorithm whose aim is twofold: First it
learns a classifier from a relatively small number of data, second it
extracts a sparse signature, {\it i.e.} a lower-dimensional
subspace carrying the information needed for the classification. In
particular the second part of the task is NP-hard, therefore we
propose a statistical-mechanics based message-passing approach. The
resulting algorithm is firstly tested on artificial data to prove its
validity, but also to elucidate possible limitations.

As an important application, we consider the classification of
gene-expression data measured in various types of cancer tissues. We
find that, despite the currently low quantity and quality of available
data (the number of available samples is much smaller than the number
of measured genes, limiting thus strongly the predictive capacities),
the algorithm performs slightly better than many state-of-the-art
approaches in bioinformatics.
\end{abstract}



\maketitle

\section{Introduction}

Extracting information from high dimensional data has become a major
challenge in biological research. The development of experimental
high-throughput techniques allows to monitor simultaneously the
behavior of genes, proteins and other cellular constituents on a
genome-wide scale. Linking the gene-expression profiles of specific
tissues to global phenotypic properties, as, {\em e.g.}, the emergence
of pathologies, is one of the most important goals of this kind of
studies.  Particular attention is paid on cancer tissues, where the
ability of classifying such tissues according to their cancer type has
immediate impact on establishing an appropriate medical treatment
\cite{Bair_Tibshirani02,
Beer_etal_NatMed_02,Bhattacharjee_PNAS_2002,Ramaswamy_Nat_Gen_2003}. Global
gene-expression profiling gives a new perspective in this direction.

In this work, we consider micro-array data, which measure the
abundance of messenger RNA as a mark of gene expression. More
precisely, the logarithm of the relative abundance compared to a
reference pattern (e.g. normal tissue or average over many tissues) is
recorded, such that positive values correspond to over-, negative to
under-expression with respect to the reference values. Each
micro-array contains the information about the expression levels of
thousands of genes, while the number of micro-arrays available for a
given problem usually ranges from some tens to few hundreds. These
 relatively few, but high-dimensional and, as a further complication,
noisy expression patterns render the classification task
computationally hard.

A promising strategy for solving this problem is that of
systematically reducing the space of the system by isolating small set
of genes that are thought to be particularly relevant for the problem.
Such a set is often called a {\it gene signature}. There are three
major reasons: (i) One may hope that the selected genes have some
biological relevance elucidating processes related to the
pathology. (ii) The reduction of the number of considered genes
improves the a priori unfavorable ratio between data dimension and
pattern number, and reduces the risk of over-fitting data, so that
better predictive capabilities are to be expected. (iii)
Custom-designed chips monitoring only the selected genes can be
constructed, reducing noise and cost of the experiments. This would
lead to an increased quality and quantity of data, and thus to better
possibilities of extracting biologically relevant information.

The problem of extracting a {\em sparse} signature from
high-dimensional datasets is ubiquitous in many different fields
ranging from computer science, combinatorial chemistry, text
processing and finally to bioinformatics , and different approach have
been proposed so far (see \cite{Guyon_Elisseeff2003,Guyon_etal_2006}
for a general introduction to the problem and
\cite{EinDor-Zuk-Domany2006} for a detailed discussion about cancer
classification based on expression patterns). The literature about
feature selection has so far concentrated around three different
strategies: (i) {\em wrapper} which utilizes learning to score
signatures according to their predictive value, (ii) {\em filters}
that fix the signature as a preprocessing step independent from the
classification strategy used in the second step. In our work we will
present a new {\em wrapper} strategy which fall into the subclass of
{\em embedded} methods where variable selection is performed in the
training process of the classifier.

Two complementary approaches can be considered to face a
classification problem: An unsupervised approach tries to suitably
regroup/cluster samples according to some inherent similarity measure
\cite{jain99data, Duda-2000-PCL}, whereas supervised approaches
exploit a (partial) labeling of the data ({\it e.g.} available
diagnoses for part of the patients) and aim at at learning a rule
which allows to label also previously unlabeled data (e.g. patients
without given diagnosis). The main problem of a supervised approach
for micro-array data is the already mentioned small number of
available labeled samples (tissues already classified) compared with
the high dimensionality (the number of genes that a priori have to be
taken into account) of each of them. On the other hand, a supervised
approach allows to capture the information that is relevant for the
sought classification, avoiding to be misled by other structures
present in the data \cite{key-martin}.

In this paper we propose a generalization of a
\textit{message-passing} based algorithm to supervised
classification. The power of the algorithm lies in its statistical
physics approach, that allows (i) to deal with the combinatorial
nature of the effect of relevant genes and (ii) to characterize the
statistical properties of the set of {\it all} possible classifiers,
weighted by their performance on the training data set, and by the
number of genes on which the classifier actually depends. The method
consists in translating the problem into a constraint satisfaction
problem (CSP), where each constraint corresponds to one classified
training pattern. This CSP is solved using {\it belief propagation}
\cite{Yedidia}, which in statistical physics is also known as the
Bethe-Peierls approximation. The final output of the algorithm is
twofold: First a gene signature is provided, and, second, the
classifier itself is given.

We test this algorithm on artificial data, we then move to data sets of
leukemia, prostate and colon cancer, used as benchmarks for several
other algorithms \cite{Dettling} and we finally consider a set of
breast cancer data obtained as the union of two different experiments
\cite{van-de-vijver,vant-veer}.

The paper is organized as follows: In Sec.~\ref{sec:model} we
introduce the problem of classification using first a
statistical-mechanics formulation, then we show the equivalence of
this formulation with the Bayesian formalism in analogy with the work
of Kabashima {\em et al.} \cite{key-kabashima}. In Sec.~\ref{sec:algo}
we give the details of the algorithm. In Sec.~\ref{sec:artdata} we
show results on artificial data and in Sec.~\ref{sec:realdata} we show
and comment the results on RNA micro-array data. Finally, in
Sec. \ref{sec:concl} we present conclusions and perspectives of our
approach.

\section{The classification problem}
\label{sec:model}

Given are $M$ patterns (micro-arrays) measuring the simultaneous
activity $x_i^\mu$ of $N$ genes, with $i=1,...,N$ and $\mu=1,...,M$.
The value of $x_i^\mu$ gives the logarithm of the ratio between the
activity of gene $i$ in pattern $\mu$ and some reference activity of
gene $i$, so positive $x_i^\mu$ correspond to over-expression,
negative to under-expression of gene $i$ in pattern $\mu$. In
addition we have an $M$-dimensional output vector $\{y^\mu\}\in
\{\pm 1\}^M$ that assigns a binary label to each pattern (e.g.
cancer vs.  normal tissue, or cancer {\tt type-1} vs.  cancer {\tt
type-2}). The full data set is thus defined as a set of input-output
pairs $D=\{(\vec x^1,y^1),\dots, (\vec x^M ,y^M) \}$.

Here we concentrate fully on binary classifications, but this
restriction can be easily overcome. A straightforward generalization
to deal with many classes is the so-called {\em one against all}
classification, where one single label is classified separately
against a set unifying all other labels. This reformulation allows
to treat a $q$-label problem via $q-1$ binary problems
\cite{Clark_Boswell_2001}.

As explained before, we aim at extracting a sparse signature out of
the set of all $N$ genes, and at establishing a functional relation
between the selected genes and the labels $y^\mu$. This task seems
infeasible in its full generality, therefore we restrict ourselves
to the simpler task: we will provide a ternary variable
$J_i\in\{0,\pm 1\}$ for each gene, which describes the influence of
gene $i$ on the output label:
\begin{equation}
J_i = \left\{
\begin{array}{rl}
-1 & {\mbox{ if over-expression of gene $i$ favors label }} y=-1 {\mbox{ and its under-expression favors label }} y=1, \\
0 & {\mbox{ if gene $i$ contains no (additional) information about label }}y, \\
1 & {\mbox{ if over-expression of gene $i$ favors label }} y=1 {\mbox{ and its under-expression favors label }} y=-1.
\end{array}
\right.
\end{equation}
We further on aim at finding variables $J_i$ such that as few entries
as possible are non-zero, forming thereby the desired sparse
signature. Note that in this scheme $J_i=0$ does not imply that
the output $y^\mu$ is independent of the input $x_i^\mu$, but that 
gene $i$ does not carry additional information about the output 
$y^\mu$ as compared to the genes in the signature (i.e., with non zero
coupling).

This classification scheme is clearly an oversimplification with
respect to bio-medical reality, where a whole range of positive and
negative interaction strengths is to be expected. On the other hand,
given the peculiar restriction posed by the limited number of
available experimental gene-expression patterns, having a simple (and
hopefully meaningful) model reduces the risk of over-fitting and
produces results which are easier to interpret.

An algorithm for binary classification based on Belief Propagation was
already proposed by Kabashima \cite{key-kabashima}, where he considers
continuous values for the couplings $ J_i$, coupled with binary
variables $c_i$ establishing the presence of the gene $i$ in the
classification task ($c_i=1$) or its absence ($c_i=0$). The algorithm
is tested only the Colon dataset and our results give a sensibly
better generalization error.

We further explore the possibility to improve our results allowing
each $J_i$ to take $q>3$ discrete values.  In all the performed
simulations the generalization ability of the algorithm remains
stationary or decreases when increasing $q$.  We thus keep the simpler
scheme considering the only possibilities $J_i \in {-1,0,1}$.


A model of the functional
relationship between input and output variables (the data set $D$) has
to be formulated to proceed. Again we aim at keeping this model as
simple as possible. We consider functions depending only on the sum of
the input variables weighted by the coupling vector. This choice,
given the Boolean nature of the output variables, results in a linear
classifier \cite{Hertz_Krogh_Palmer_1991}, {\em i.e.}~a perceptron,
\begin{equation}
\label{eq:model}
y^\mu=\mathrm{sign}\left(\sum_{i=1}^{N}J_{i}x_{i}^\mu+ \tau\right)
\end{equation}
where the function $\mathrm{sign}\left(x\right)$ gives the sign of the
input $x$, and $\mathrm{sign}(0)$ is set to +1.

The full classification problem reduces thus to the inference of a
coupling vector $\vec J=(J_1,...,J_N)$.

\subsection{Inference as a constraint satisfaction problem}

The coupling vector $\vec J$ and the threshold $\tau$ are the free
parameters of the problem.  Following the usual strategy in
statistical mechanics \cite{Braunstein_Zec_PRL2005}, we
can define a cost function (or Hamiltonian) that, given the data set
$D$, counts the number of patterns contradicting our threshold model
as a function of the coupling vector $\vec J$ (and of the threshold
$\tau$; this dependence will be taken as implicit from now on):
\begin{equation}
\label{eq:H0}
  {\cal H}_0(\vec J\ ) =\sum_{\mu=1}^{M}
  \Theta\left\{
  -y^{\mu} \left(
  \sum_{i=1}^{N}J_{i}x_{i}^{\mu} + \tau
  \right) \right\}\ ,
\end{equation}
with $\Theta$ being the Heaviside step function. This cost function does not
include any dependence on the sparsity of the coupling vector $\vec
J$, so, to obtain a vector with as many zero entries as possible, we
have to add an external pressure to our system. From the point of
view of a model Hamiltonian, this can be obtained by an {\it
external field} $\tilde h$ (or chemical potential) being coupled to
the number of non-zero entries in $\vec J$, i.e. to
\begin{equation}
N_\mathrm{eff}(\vec{J})=\sum_{i=1}^{N}|J_{i}|\ ,
\end{equation}
The complete Hamiltonian for our system thus reads
\begin{equation}
\label{eq:Ham}
{\cal H}(\vec{J}) = {\cal H}_0(\vec{J}) + \tilde h N_\mathrm{eff}(\vec{J})\ .
\end{equation}
Searching for the minimum of this cost function is the analogous
of solving a zero temperature problem in statistical mechanics.

This would be the correct procedure if the data were completely clean from noise
and perfectly linearly separable.
When this is not the case, imposing the correct classification for all
the patterns could lead to an improper selection of the coupling vector $\vec J$
and consequently to a poor prediction ability.

Keeping this in mind, we consider  a ``finite temperature problem'' instead,
where also solutions that leave unsatisfied some clauses are allowed. We
see that in many cases these are the solutions that give the better predictions
on new data.
At this aim, we introduce a formal inverse temperature $\beta$ and the
related Gibbs measure
\begin{equation}
\label{eq:gibbs}
P_\mathrm{Gibbs}(\vec J) \propto \exp(-\beta {\cal H}_0 - h N_\mathrm{eff})
\end{equation}
with $h=\beta\tilde h$. The values of $\beta$ and $h$ set the relative
importance of satisfying the constraints given by the patterns versus
the sparsity of the coupling vector.  Large $\beta$ enforces
satisfaction of the constraints, large $h$ favors many zero-elements
in $\vec J$. At the moment these two parameters are free model
parameters and we describe later on a strategy to fix them for
specific data sets $D$.

However, our approach is not seeking for one single configuration
that minimizes the cost function; the objective is to characterize
the statistical properties of {\it all} low-cost coupling vectors as
weighted by $P_\mathrm{Gibbs}(\vec J)$.
 Once known these statistical properties, a subsequent problem
 is to characterize a proper classifier. We will discuss this point
 in the next subsection, where, to clarify the relation between our statistical
mechanics and the Bayesian approach, we reformulate the problem
using the point of view of the latter.

\subsection{A Bayesian point of view}

The statistical-mechanics approach outlined so far can be
reinterpreted in terms of the following Bayesian inference scheme
\cite{McKayBook}. As a first step, we define a model that describes
the likelihood $P( \{y^\mu\}_{\mu=1,...,M} \,|\,\vec J,\{\vec
x^\mu\}_{\mu=1,...,M} )$ of a labeling $\{y^\mu\}_{\mu=1,...,M}$
given a parameter vector $\vec J$ and the expression data $\{\vec
x^\mu\}_{\mu=1,...,M}$. As usually we assume different data points
to be generated independently under the model, i.e.
\begin{equation}
\label{eq:bayesmodel}
P\left( \{y^\mu\}_{\mu=1,...,M} \,|\,\vec J,\{\vec x^\mu\}_{\mu=1,...,M} \right) =
\prod_{\mu=1}^M P(y^\mu\, | \, \vec J, \vec x^\mu ) \ ,
\end{equation}
with the likelihood of a single label being defined coherently with the
previous subsection,
\begin{equation}
P(y^\mu\,|\,\vec J, \vec x^\mu) = \frac 1{1+e^{-\beta}}\
\exp\left\{ -\beta \Theta\left( -y ^\mu \left[\sum_{i=1}^N J_i x_i^\mu \right]
\right)  \right\}\ .
\end{equation}
Further on we define a prior on the space of parameters which favors
again sparse vectors: $\Pi( \vec J ) \propto \exp \{ - h
N_\mathrm{eff}(\vec J)
\}(\delta(J_i;0)+\delta(J_i;-1)+\delta(J_i;-1))$ and where we used the
Kronecker delta function to enforce that $J_i\in\{0,\pm 1\}$.

A straightforward application on Bayes theorem allows us to derive the
posterior probability of $\vec J$ given the knowledge of the data-set
$D$ as:
\begin{equation}
\label{eq:post}
P(\vec J\,|\, D) \propto
P( \{y^\mu\} \,|\,\vec J,\{\vec x^\mu\} )\  \Pi( \vec J ) =
\prod_{\mu=1}^M P(y^\mu\,| \, \vec J, \vec x^\mu )\ \Pi( \vec J )
\end{equation}
which is the analogous of Eq.~(\ref{eq:gibbs}).

Let us now see how the knowledge of this posterior probability can be
used in a given case of classification. Imagine that we have
experimental access to a data-set $D$ of $M$ gene-array measurements
$D = \{ (\vec x^1,y^1),\dots, (\vec x^M ,y^M) \}$. Now a new
expression measure $\vec x^0$ becomes available, but we do not know
whether the experimental sample comes from a cancerous tissue or not,
{\em i.e.} we do not know its annotation label $y^0$. We can, however,
compute the conditional probability of $y^0$ given the knowledge of
all experimental measurements in $D$ and the new expression profile
$\vec x^0$. Applying the sum rule:
\begin{equation}
\label{eq:sumrule}
P( y^0 | D , \vec x^0 )= \sum_{\left\{ \vec{J}\right\}} P(y^0,\vec
J\,\,|\,\, D, \vec x^0) =
\sum_{\left\{ \vec{J}\right\}} P(
y^0 | \vec J ,\vec x^0 ) P(\vec J\, |\, D )\,\,\,\,\,,
\end{equation}
we are ready to establish two probabilistic rule to assign the label
$y^0$ to a sample $\vec{x}^0$ on the basis of the posterior
probability $P(\vec J \,\,|\,\, D)$.  To simplify the notation, let
us define the quantity:
\begin{equation}
\label{eq:H}
H \equiv \sum_{i=1}^N J_{i}x_i^0
\end{equation}
We can then classify the new pattern $\vec x\,^0$ according to the
Bayesian rule (\ref{eq:sumrule}). If we assume that (\ref{eq:H}) is a
sum of random independent variables, we get that the field $H$ is
approximately a Gaussian distributed random variable. In this way we
can relate the probability distribution of $H$ with the conditional
probability of the output $y^0$:
\begin{eqnarray}
  \langle y^0 \rangle  &=&   P\left(y^0 = 1 | D ,  \vec{x}^0 \right) -
  P\left(y^0 = - 1 | D , \vec{x}^0 \right) = \sum_{\{ \vec{J}\}}
  \left(
  \frac{e^{-\beta\Theta(-\sum_{i=1}^N J_i x_i^0)}}{1+e^{-\beta}}
  -\frac{e^{-\beta\Theta(\sum_{i=1}^N J_i x_i^0)}}{1+e^{-\beta}}
  \right)
  P(\vec{J}|D)\nonumber\\ 
  &\simeq&
  \int\left( 
  \frac{e^{-\beta\Theta(-H)}}{1+e^{-\beta}}
  -\frac{e^{-\beta\Theta(H)}}{1+e^{-\beta}}
  \right)G_{ \langle H
    \rangle , \langle \Delta H ^2\rangle  } (H) dH = 
  \frac{1-e^{-\beta}}{1+e^{-\beta}} \int dH\ G_{ \langle H
    \rangle , \langle \Delta H ^2\rangle  } (H)\ \mathrm{sign}(H)
\end{eqnarray}
where $G_{m,\sigma}(H)$ is a Gaussian distribution with mean $m$
and variance $\sigma$. In our case these two parameters are given by:
\begin{eqnarray}
  \langle H \rangle &=&\sum_{ i=1}^N {\langle J}_{i} \rangle\
  x_i^0\\ \langle \Delta H^2\rangle \equiv \langle H^2 \rangle- \langle
  H {\rangle}^{2} &=& \sum_{ i=1}^N\left({\langle J}_{i}^{2}\rangle
  -{\langle J}_{i}\rangle ^{2}\right)\ (x_i^0)^{2}\ .
\end{eqnarray}
The averages $\langle \cdot \rangle$ in the previous equation are
taken over the posterior probability distribution $P( \vec
J\,|\,D)$. The choice of the label $y^0$ can be simply given by the
maximum posterior probability criterion:
\begin{equation}
  y^0= \mathrm{sign}(\langle y_0\rangle) = \mathrm{sign}(\langle H\rangle)\ .
\end{equation}

It is worth pointing out that for computing both $\langle y^0\rangle$
and $\langle H \rangle$ we do not need the knowledge of the whole
posterior probability. The knowledge of the single-coupling marginal
posterior probabilities $P_i (J_i \,|\, D ) \equiv
\sum_{\{J_l\}_{l\neq i }} P( \vec J \,| \, D ) $ is sufficient. In
order to lighten notations, we indicate marginal posterior
probabilities simply by $P_i(J_i)$, dropping the explicit dependence
on the data set $D$.

In the following section we introduce an efficient algorithm for
estimating these quantities.

\section{The algorithm}
\label{sec:algo}
 The message-passing strategy introduced in the
following allows to efficiently estimate marginal probability
distributions for single entries of the coupling vector. Given the
Gibbs measure (\ref{eq:gibbs}) one could in principle compute the
marginal probability distribution of variable $i$ using the standard
definition:
\begin{equation}
P_i(J_i)\equiv\sum_{\{J_j\}_{j \neq i}}P_\mathrm{Gibbs}(\vec J ) \ .
\end{equation}
Unfortunately this strategy involves a sum over $3^{N-1}$ terms and
becomes computationally infeasible already for $N$ larger then 30, as
compared to thousands of gene-expression values measured in each
micro-array experiment. In the next section we will explain how to
overcome this difficulty by using the Bethe-Peierls approximation.

Note that the marginal distributions $P_i(J_i)$ contain valuable
information for the extraction of a sparse signature. Genes $i$ with
a high probability of having a non-zero coupling $J_i$ even for
large $h$ have to be included in such a signature, they are likely
to carry crucial information about the output label. Genes $i$ with
high probability $P_i(J_i=0)$ on the other hand can be excluded from
most signatures, so their information content is either low or
already provided by other genes.

\begin{figure}
\centerline{\includegraphics[width=7cm]{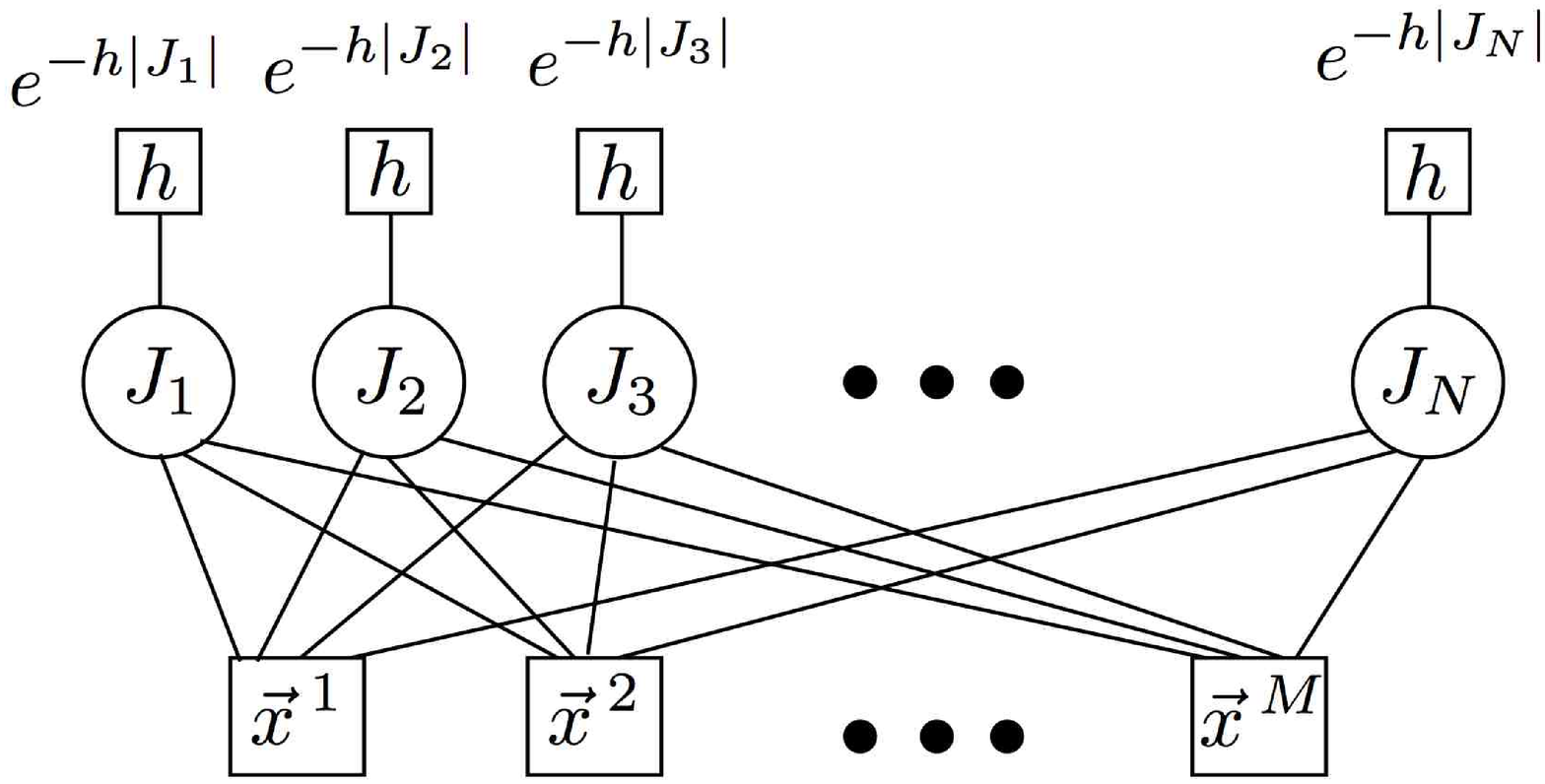}}
\caption{Factor graph representation of the problem. Circles are
  variables nodes, corresponding to single coupling variables $J_i$,
  rectangles represent factor nodes, the lower ones corresponding to
  data-given constraints, the upper (trivial) ones to the external
  diluting field. Links indicate functional dependence.}
\label{fig:factor_graph}
\end{figure}

\subsection{The message passing approach}
\label{sec:mp} The \textit{belief propagation} (BP) method, also
known as Bethe-Peierls approximation, or cavity approach, in
statistical physics, is exact on tree-shaped graphical models. As an
approximate tool, it was therefore mostly used on sparse graphs
\cite{Yedidia,SumProd,BMZ}, where the influence of loops is expected
to be not very strong. More recently, several applications to dense
graphs have been successfully proposed
\cite{Kabashima2007,key-kabashima,key-Bickson,Braunstein_Zec_PRL2005,Frey,LeoneSumedhaWeigt}.
Our problem can in fact be considered as a graphical problem over a
fully connected bipartite {\it factor graph},
cf.~Fig.~\ref{fig:factor_graph}, with $N$ vertices (variable nodes)
representing the $N$ variables $J_i$, and $M$ vertices (factor nodes)
representing the $M$ constraints (\ref{eq:model}). All pairs of these
two vertex types are connected, since each constraint depends on all
variables. In addition there are local fields on variables
corresponding to the diluting term in Hamiltonian (\ref{eq:Ham}).

The BP algorithm provides a strategy for estimating marginal
probability distributions. It works via iterative updates of
messages, which are exchanged between variable and factor nodes. Let
$\mu$ be one of the factor nodes and $i$ one of the variable nodes.
We can introduce the following messages, which travel according to
the direction indicated by the indices:
\begin{itemize}
\item $\rho_{\mu \rightarrow i } (J_i)$ describes a weight imposed by
  constraint $\mu$ on the value $J_i$ of the coupling of variable $i$.
\item $P_{i \rightarrow \mu}(J_i)$ is the probability that variable
  $i$ takes value $J_i$ in the absence of the constraint set by factor
  node $\mu$.
\end{itemize}

The above-defined quantities satisfy the following set of
self-consistent BP equations:
\begin{eqnarray}
\label{eq:bp1}
\rho_{\mu\rightarrow i}\left(J_{i}\right)
& = & \sum_{\left\{ J_{j}\right\} _{j\neq i}}
\exp\left\{-\beta\Theta\left(-y^{\mu}\sum_{s=1}^{N}J_{s}x_{s}^{\mu}\right)\right\}\prod_{j\neq
  i}P_{j\rightarrow\mu}\left(J_{j}\right)
\\
P_{i\rightarrow\mu}\left(J_{i}\right)
&=& C_{i \to \mu} e^{-h|J_{i}|}\prod_{\nu\neq\mu}\rho_{\nu\rightarrow i}
\left(J_{i}\right)
\label{eq:bp2}
\end{eqnarray}
where $C_{i \to \mu}$ is a normalization constant. A more detailed
derivation of the equations in the case of the perceptron without
dilution can be found in \cite{Braunstein_Zec_PRL2005}, and
in \cite{key-martin} in the case of dilution.

In the algorithm both types of messages are initialized randomly, and
the iteration proceeds via a random sequential update scheme. The
algorithm stops when the convergence is reached, i.e.  when the
difference between each message at time $t$ and the corresponding one
at time $t-1$ is less than a predefined threshold ($10^{-12}$ in all
our simulations). The number of needed iterations depends on the
particular problem and on the values settled for the parameters
$\beta$ and $h$. However, we do not exceed a hundred of iterations in
the simulations reported above.

Once all messages are evaluated, the desired marginal probability is
given by the messages sent from {\it all} factor nodes and by the
diluting field,
\begin{equation}
  P_i( J_i ) = C_i  e^{-h|J_{i}|}\prod_{\nu}\rho_{\nu\rightarrow i}\left(J_{i}\right)
  \label{eq:bpsitemarg}
\end{equation}
where the $C_i$ are normalization constants which can be easily
determined by tracing the unnormalized expression over the three
values $J_i=0,\pm 1$.

Note that in the definition of messages (\ref{eq:bp1}) we still have a
sum over $3^{N-1}$ configurations, but they enter in the expression
only through a linear combination, which enables us to use again a
Gaussian approximation \cite{Braunstein_Zec_PRL2005}. The sum over 
couplings is replaced by a single Gaussian integral, which can be 
performed explicitly:
\begin{eqnarray}
\label{eq:gaussianapp}
\rho_{\mu\rightarrow i}\left(J_{i}\right) &=&
\int_{-\infty}^{+\infty} \frac{dz_{\mu\rightarrow i}}{\sqrt{2\pi\sigma_{\mu\rightarrow i}^{2}}}
\exp\left\{-\frac{\left(z_{\mu\rightarrow i}-\bar{z}_{\mu\rightarrow i}\right)^{2}}
{2\sigma_{\mu\rightarrow i}^{2}}\right\}
\exp\left\{-\beta\Theta\left(-y^{\mu}\left(z_{\mu\rightarrow i}+J_{i}x_{i}^{\mu}\right)\right)\right\}\\
&=& \frac{1+e^{-\beta}}2+ \frac{1-e^{-\beta}}2\ {\rm erf}\left\{ y^{\mu}
\frac{\bar{z}_{\mu\rightarrow i}+J_{i}x_{i}^{\mu}}{\sqrt{2\sigma_{\mu\rightarrow i}^{2}}}\right\}
\end{eqnarray}
where $z_{\mu\rightarrow i}$ is a Gaussian variable with mean and
variance,
\begin{eqnarray} \bar z_{\mu\rightarrow i} & = &
  \sum_{j\neq i}
  \langle J_j\rangle_{j\to\mu}\ x_j^\mu\nonumber \\
  \sigma_{\mu\rightarrow i}^2 & = &
  \sum_{j\neq i} \left({\langle J}_{j}^{2}\rangle_{j\to\mu}\
  -{\langle J}_{j}\rangle_{j\to\mu}^{2}\right) {x_j^\mu}^2\ .
\label{eq:siggauss}
\end{eqnarray}
The averages $\langle \cdot \rangle_{j\to\mu}$ are performed over
messages $P_{j\to\mu}(J_j)$. In this way the complexity of
Eq.~(\ref{eq:bp1}) is reduced from ${\cal O}(3^N)$ to ${\cal O}(N)$
and that of the overall iteration to ${\cal O}(M N)$ 
(The apparent complexity ${\cal O}(MN^2)$ of updating $MN$
messages in time ${\cal O}(N)$ can be reduced to ${\cal O}(MN^2)$
by a simple trick: The sums in Eqs.~(\ref{eq:siggauss}) can be calculated
over all $j$ once for each $\mu$, so only the contribution of $i$ has to 
be removed in the update of $\rho_{\mu\to i}$ for each $i$. This allows 
to make the single update step in constant time). A precise
estimate of the overall complexity of the algorithm would require
to control the scaling of the number of iterations needed for 
convergence. A theoretical analysis of BP convergence times in
a general setting (including the perceptron case) remains elusive. 
Some recent progress for the simpler matching problem can be found in
\cite{Zec_matching}.

The BP equations become feasible even for very large $N$ and $M$, and
can therefore be applied to biological high-throughput data sets. Note
that, even if the central limit theorem is meant to be valid in the
limit of $N\rightarrow \infty$, in practice it works very well already
for $N=4$ (where the exact computation is clearly feasible).

The Bethe entropy \cite{Yedidia} can be computed from marginals and
messages:
\begin{eqnarray}
S_{\rm Bethe} &=& - \sum_{\mu=1}^M \sum_{\vec J} P_\mu(\vec J) \ln P_\mu (\vec J) +
(M-1) \sum_{i=1}^N \sum_{J_i = -1,0,1} P_i(J_i) \ln P_i(J_i)\nonumber\\
&=&-\sum_{\mu=1}^M \ln C^\mu -  \sum_{\mu =1 }^M C^\mu \sum_{i =1 }^N
\sum_{J_i = -1,0,1} \rho_{\mu\rightarrow i}(J_i)P_{i\rightarrow\mu}(J_i)
\ln P_{i\rightarrow\mu}(J_i) + \beta  \sum_{\mu =1 }^M C^\mu E_{\mu} \nonumber \\
&& +(M-1) \sum_{i=1}^N \sum_{J_i = -1,0,1} P_i(J_i) \ln P_i(J_i)
\end{eqnarray}
In the first line, we have used the distribution
\begin{equation}
  P_\mu( \vec J ) = C^\mu
\exp\left\{-\beta\Theta\left(-y^{\mu}\sum_{s=1}^{N}J_{s}x_{s}^{\mu}\right)\right\}
\prod_{j=1}^N P_{j \to \mu}(J_j)
\end{equation}
which describes the influence of a single factor node by conserved
marginal distributions.  In the second line, we use the corresponding
single sample energy $c^\mu E_{\mu}$ defined as
\begin{eqnarray}
E_{\mu}&=& \sum_{\left\{ J_{j}\right\} _{j\neq i}}
\Theta\left(-y^{\mu}\sum_{s=1}^{N}J_{s}x_{s}^{\mu}\right)
\exp\left\{-\beta\Theta\left(-y^{\mu}\sum_{s=1}^{N}J_{s}x_{s}^{\mu}\right)\right\}
\prod_{j\neq i}P_{j\rightarrow\mu}\left(J_{j}\right)\nonumber\\
& = & \frac{e^{-\beta}}2 \left[1- {\rm erf}\left\{y^{\mu} \frac{\bar{z}_{\mu\rightarrow i}
+J_{i}x_{i}^{\mu}}{\sqrt{2\sigma_{\mu\rightarrow i}^{2}}}\right\}\right]\ .
\end{eqnarray}
In writing the last expression we have used again the Gaussian
approximation as in Eqs.~(\ref{eq:gaussianapp}-\ref{eq:siggauss}).

Up to this point, the BP equations still depend on two undetermined
parameters, namely the inverse temperature $\beta$ coupled to the
data-given constraints, and the diluting field $h$. To implement the
algorithm we have to define a strategy for fixing these free
parameters:
\begin{itemize}
\item The diluting field $h$ is the conjugate variable of the number
  of effective link $N_{\mathrm{eff}}(\vec J)$, so we can equivalently
  fix one of the two quantities. We decided to fix the number of
  effective links, and thus the size of the searched gene signature,
  and to choose $h$ accordingly. To find the correct value of $h$ we
  apply a cooling procedure, where after each interaction of the BP
  equations step we increase (resp. decrease) $h$ depending on whether
  the effective number of link is higher (resp. lower) than the
  desired value. Since the true number of relevant genes is an unknown
  quantity, the chosen value for $N_{\mathrm{eff}}(\vec J)$ is,
  however, a free parameter. Practically, since the algorithm finds marginal
  probabilities and not a single configuration, we define the number
  of relevant links via its thermodynamic average:
  \begin{equation}
    \langle N_{\mathrm{eff}}\rangle =\sum_{i=1}^N [1-P_i(0)]\ .
  \end{equation}
  Comparing results for different values of $\langle N_{\mathrm{eff}} \rangle$
  we see that the algorithm is fairly robust, as will be seen in the
  following sections.
\item The inverse temperature $\beta$ is again fixed by a cooling
  procedure starting from a low value and increasing it until one of
  the following two conditions is met: (i) the energy reaches a small
  enough value ($\beta\to\infty$ formally corresponds to zero energy),
  (ii) the entropy goes to zero (a signal for a freezing transition
  into a non-extensive number of coupling configurations at finite
  temperature).  In this last case we use the marginals computed at
  the zero-entropy temperature.
\end{itemize}

The diluting field $h$ drives the system toward solutions of the
desired average number of effective coupling $\langle
N_{\mathrm{eff}}\rangle$.  This is not yet enough for determining
explicitly one signature of the desired size, since results are still
of probabilistic nature.
If we want to select, and use in our algorithm, only a desired number of genes,
we have to couple the BP algorithm with
the {\em decimation} procedure presented in
\cite{ScienceZec,Zec_Mezard_2002}:
\begin{enumerate}
\item Random initialization of messages.
\item Convergence of the BP Eqs.~(\ref{eq:bp1}) and (\ref{eq:bp2}) at
  the correctly self-tuned values of $h$ and $\beta$.
\item Computation of marginal probability distributions using Eq.
  (\ref{eq:bpsitemarg}).
\item Setting to zero the coupling variable $J_i$ having the largest
  weight in zero ({\em i.e.}, set $P_i(J_i=0 ) = 1$ for this variable).
\item GOTO 2. Repeat until $N- N_{\mathrm{eff}}$ classification variables are
  set to zero.
\end{enumerate}
Practically, this procedure turned out to be computationally much
too expensive, so we opted for a faster variant of Step 4 of the
decimation procedure: after each convergence step of the BP
equations, we rank genes according to $P_i(0)$, and we set to zero
an extensive part of couplings at the top of this ranking. The same
procedure is iterated until we reach the desired number of non zero
weights $J_i$.

\subsection{A note on the centroids algorithm}

It is interesting to compare the results of the BP algorithm with
simpler, but widely applied techniques. The centroids algorithm is
based on the notion of distance (Euclidean distance in $N$-dimensional
space in this case) of a given pattern from the centers of mass of the
two sets of patterns with annotations $y^\mu = 1$ or $-1$. The
algorithm works in the following way:
\begin{itemize}
\item Let $B_{z} \equiv \{ \mu \,| \, y^\mu = z \}$, and $M_z=|B_z|$
  the number of patterns with label $z$. Here we consider $z=\pm 1$
  only, but the algorithm works equivalently for multiple classes. We
  compute the centers of mass (or centroids) $\vec{c}_{z}$ of the
  expression data with labels $y=z$:
\begin{equation}
\vec{c}_{z} = \frac 1{M_z} \sum_{\mu\in B_{z}} \vec{x}^{\mu}
\end{equation}
\item We assign the label $y^0=z$ to a new sample $\vec x\,^0$ if its
  distance from $\vec{c}_{z}$ is smaller than its distance from all
  other centroids,
  \begin{equation}
    y^0 = \underset{z}{\operatorname{argmin}}( || \vec x\,^0 -  \vec c_{z} || )\ ,
  \end{equation}
where $ || . ||$ can be the Euclidean distance in $\mathbb{R}^N$ or, depending on the
problem, any other meaningful notion of distance.
\end{itemize}
In order to take into account that only a subset of the genes is
relevant (sparse signature), we rank genes $i=1,...,N$ according to
the absolute value of their Pearson correlation $C_i$ with the output:
\begin{equation}
C_i=\frac{ \frac 1M \sum_{\mu=1}^M x_i^{\mu} y^{\mu}- \left( \frac 1M \sum_{\mu=1}^M
y^{\mu} \right) \left( \frac 1M \sum_{\mu=1}^M x_i^{\mu} \right)
}{\sqrt{1-\left( \frac 1M \sum_{\mu=1}^M y^{\mu}
\right)^2}\sqrt{ \frac 1M \sum_{\mu=1}^M {x_i^{\mu}}^2
-\left( \frac 1M \sum_{\mu=1}^M x_i^{\mu} \right)^2}   } \ .
\label{pearson}
\end{equation}
The $s$ highest scoring genes are selected as the signature, and the
distance of the new pattern $\vec x\,^0$ from the centroids is
computed taking into account only these genes, {\em i.e.} the full
problem is projected onto the $s$-dimensional subspace of all
genes. Since the signature size $s$ yielding the best classification
is not known a priori, one has to consider the performance of the
algorithm as a function of it.

\begin{figure}[htb]
\includegraphics[width=0.32\columnwidth]{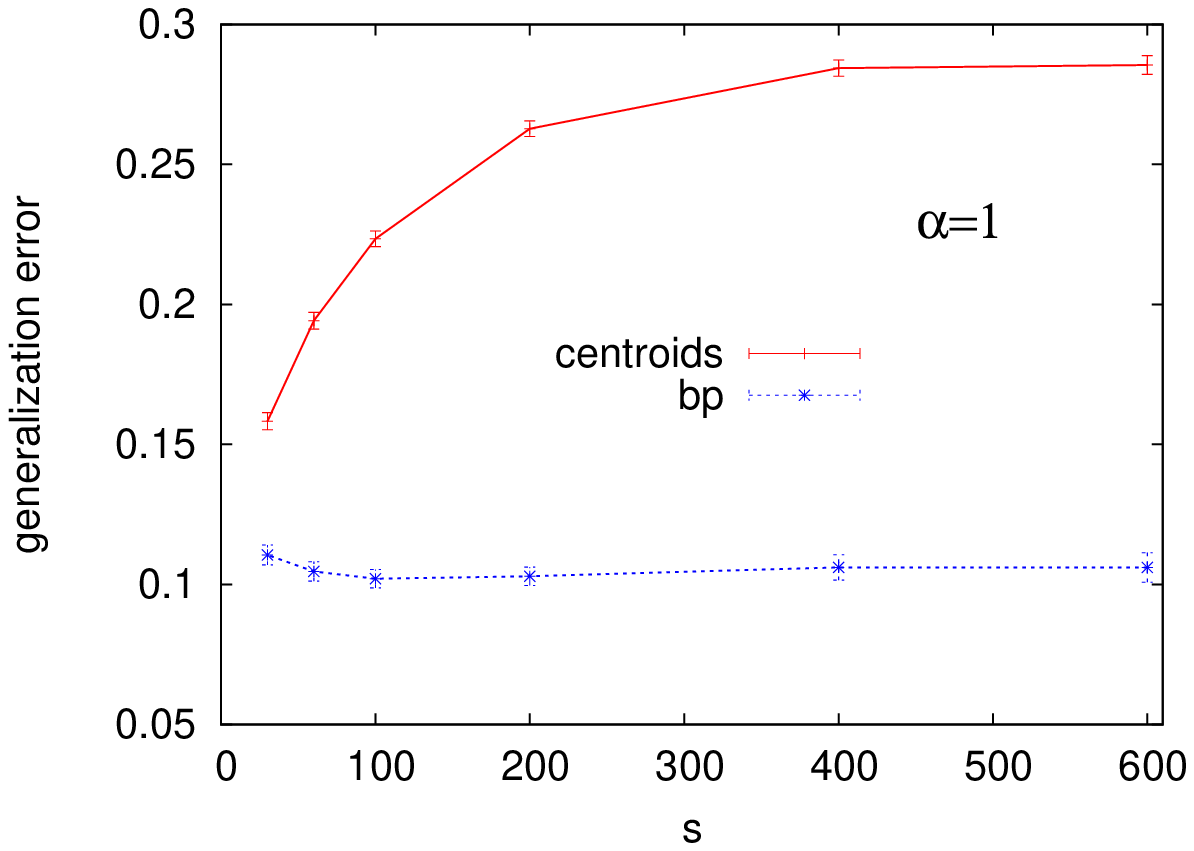}
\includegraphics[width=0.32\columnwidth]{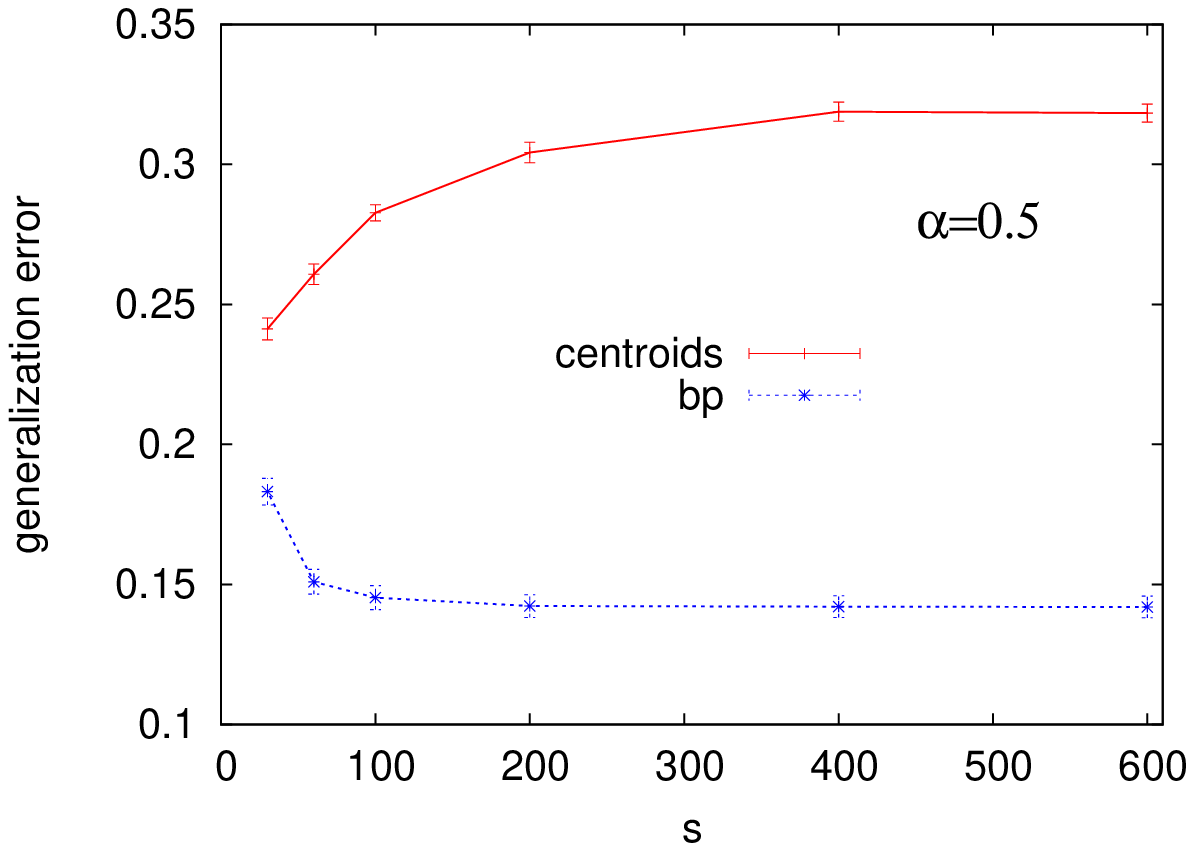}
\includegraphics[width=0.32\columnwidth]{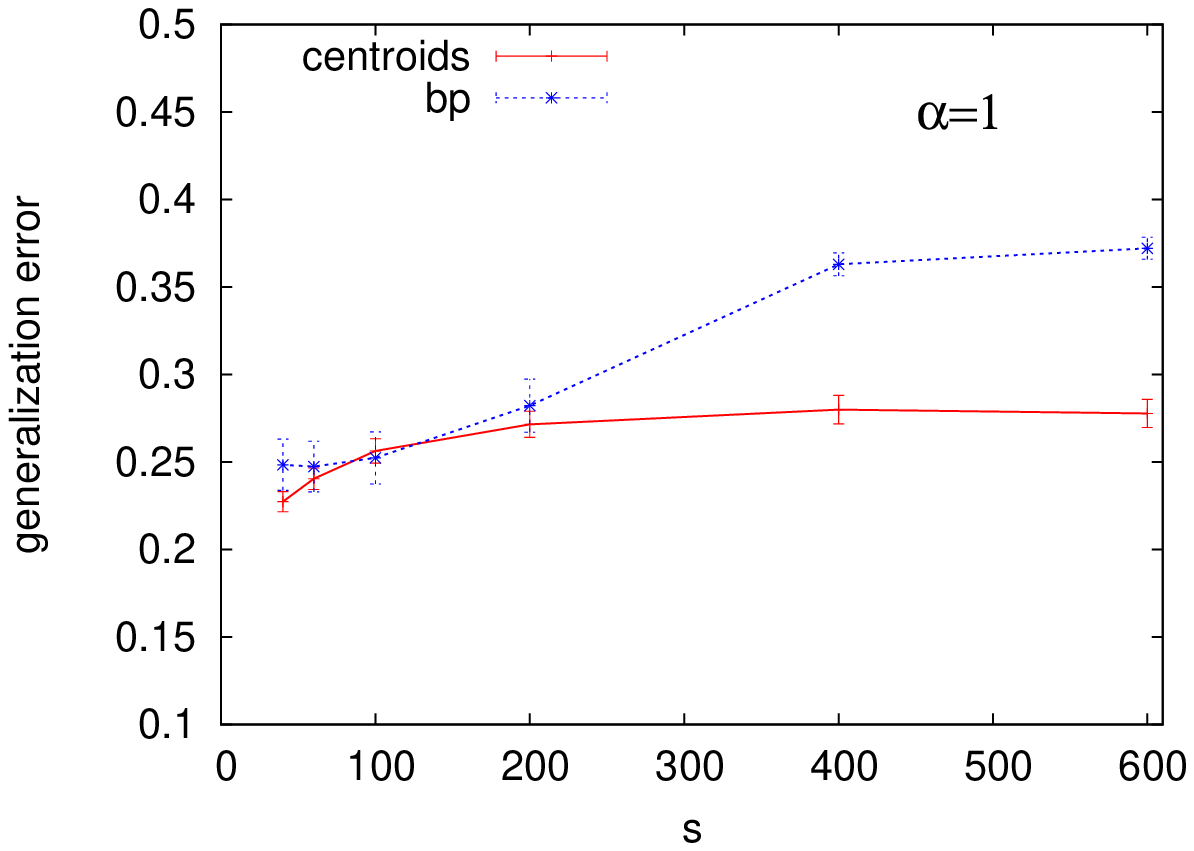}
\caption{Generalization error as a function of the signature size $s$
  for artificial data generated according to
  Eqs.~(\ref{Pj0},\ref{eq:sign}) with $N=600$ and $k_1=k_2=0.025$
  (left and central panel) and according to the centroids rule with $k=6$ (right panel). In
  the left and right panel we have $M_\mathrm{training}=600$ ($\alpha=1$), in
  the central panel $M_\mathrm{training}=300$ ($\alpha=0.5$). In all the
  cases $M_\mathrm{test}=300$.  The curves are averages over 50
  different realizations of the training and test sets.}
\label{generr_art}
\end{figure}

\begin{figure}[htb]
\includegraphics[width=0.32\columnwidth]{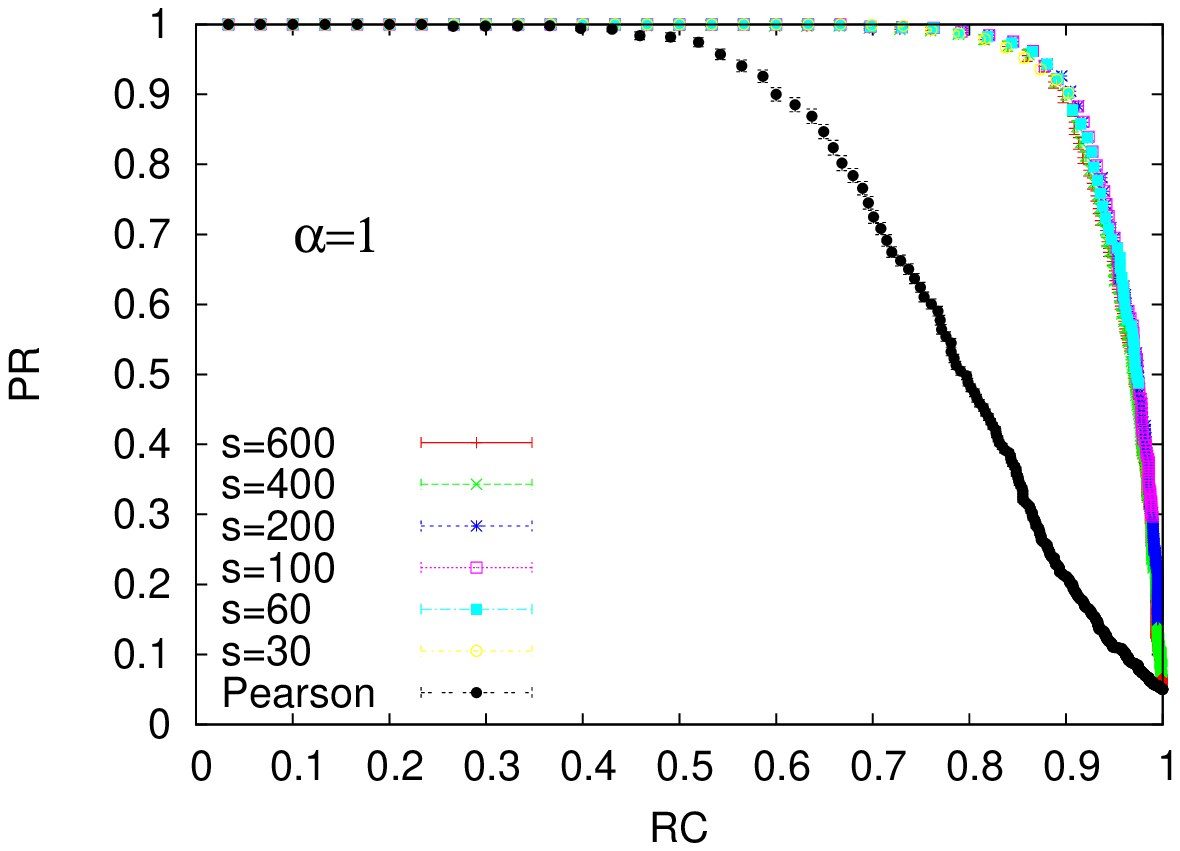}
\includegraphics[width=0.32\columnwidth]{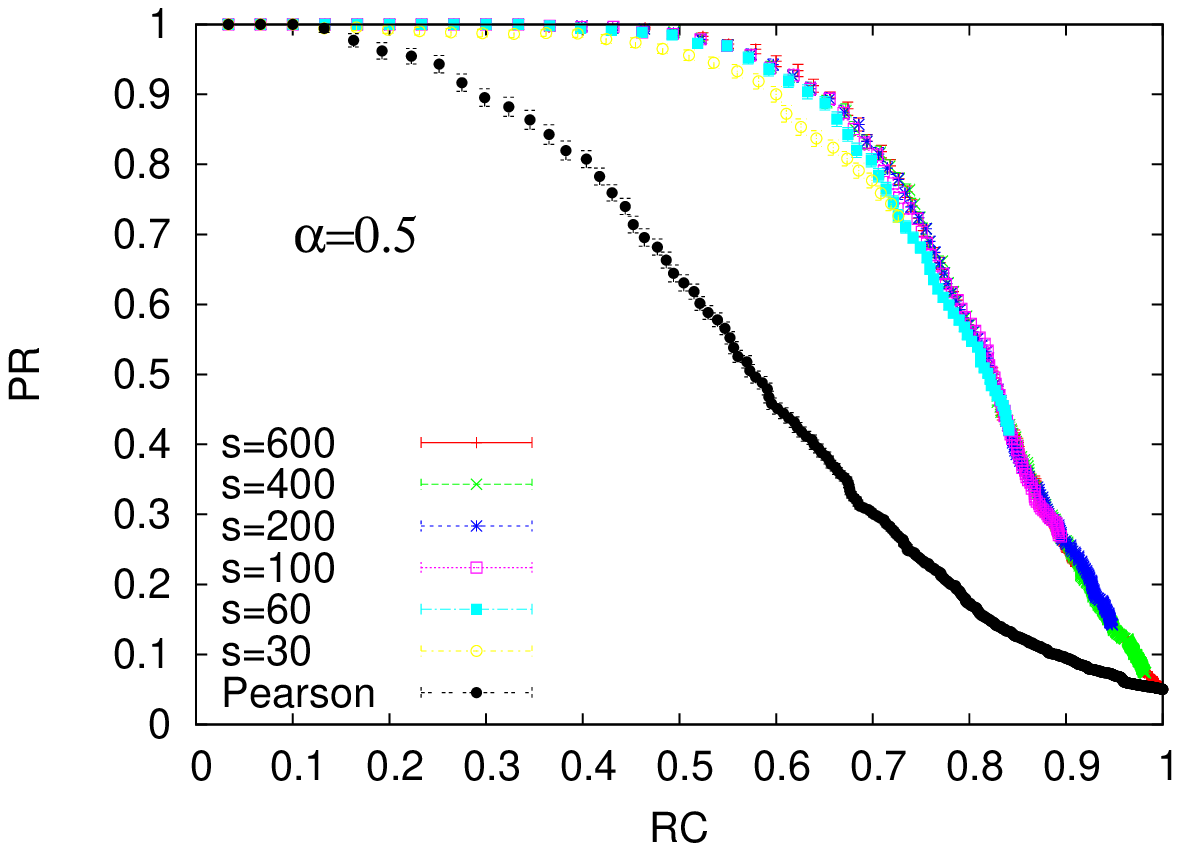}
\includegraphics[width=0.32\columnwidth]{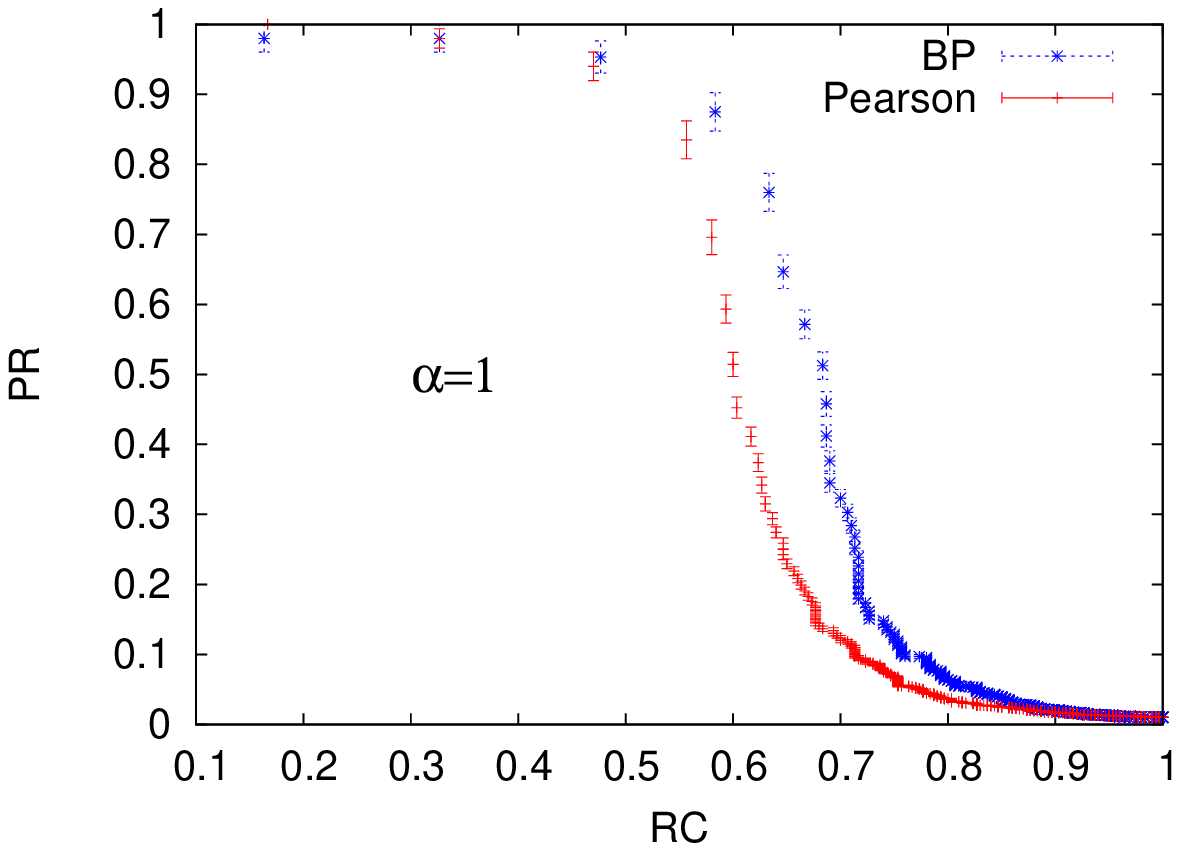}
\caption{Precision-versus-recall curves for the teacher-student
  scenario. Results for the perceptron-like dataset in the case
  $\alpha=M_\mathrm{training}/N=1$ (left panel) and
  $\alpha=M_\mathrm{training}/N=0.5$ (central panel), and for the centroids-like dataset
  for $\alpha=M_\mathrm{training}/N=1$ (right panel).
  We display curves
  obtained from the BP algorithm with decimation (see text) for
  different values of the signature $s$. The curves derived from the
  Pearson correlation ranking used in the centroids algorithm are
  significantly lower than the BP ones in the perceptron-like dataset case. The same recall
  leads to lower precision in this case. In the centroids-like dataset case (displayed $s=600$),
  the curves for the two algorithms are comparable but BP continues to perform slightly better.
  Performance goes slightly down for smaller $s$ but stays comparable.
  In the left and right panel we have $M_\mathrm{training}=600$ ($\alpha=1$), in
  the central panel $M_\mathrm{training}=300$ ($\alpha=0.5$). In all the
  cases $M_\mathrm{test}=300$.  The curves are averages over 50
  different realizations of the training and test sets.}
\label{roc_art}
\end{figure}

\section{Test on artificially generated data}
\label{sec:artdata}

Before running the algorithm on micro-array data, it is useful to test
it on artificial data generated by a controlled input-output
relation.
 In the simulations reported here the $x_i^\mu$ are drawn
independently from a normal distribution.

In order to compare the performance of the BP and the centroids algorithms,
we label the random generated patterns according to two different rules, reflecting
the perceptron and the centroids idea respectively.

\subsection{Description of the datasets.}

In the first series of simulations,
we draw the couplings $J_i^0$
from the following
distribution:
\begin{equation}
P_{1}(J_{i}^{0})=(1-k_{1}-k_{2})\delta_{J_{i}^{0},0}
+\frac{k_{1}}{2}[\delta_{J_{i}^{0},1}+\delta_{J_{i}^{0},-1}]
+\frac{k_{2}}{2}[\delta_{J_{i}^{0},2}+\delta_{J_{i}^{0},-2}]\ .
\label{Pj0}
\end{equation}
We consider the case $k_1+k_2 \ll 1$ in order to have a sparse gene
signature, the expectation value of $N_{\rm eff}$ (defined as the
number of non-zero entries of $\vec J\,^0$) equals
$(k_1+k_2)N$. Labels are determined according to a rule similar to the
one used for BP inference,
\begin{equation}
\label{eq:sign}
 y^{\mu}=\mathrm{sign}\left(\sum_{i=1}^{N}J_{i}^{0}x_{i}^{\mu}\right)\ .
\end{equation}

If $k_2=0$ only $J_i^0=0,\pm 1$ exist, values $\pm 2$ are excluded. In
this case the data are feasible under model if Eq.~\ref{eq:model}, and
data can be learned at zero energy (no wrongly assigned labels by the
inferred vector $\vec J$).  The theoretical analysis presented in
\cite{key-martin} shows a phase-transition line in the plane
$(\alpha,h)$, with $\alpha=M/N$. One phase, at low $\alpha$ and $h$,
is paramagnetic. It perfectly memorizes the input-output relation
given by the data, but there is no correspondence between the
exponential number of possible coupling vectors and the data
generating rule, the predictive properties for a new pattern $\vec
x\,^0$ are poor. For higher $h$ and/or $\alpha$, the solution
discontinuously jumps to a perfect retrieval of the input-output
association vector $\vec J\,^0$. A particularly important point is
that at sufficiently high diluting field $h$ perfect inference is
possible at $\alpha$-values much lower than the critical threshold
$\alpha_c$ for perfect inference at $h=0$.

In the case $k_{1,2} \neq 0$ the data are infeasible, since the
structure of the data generator is richer than the one used for
inference. Couplings of the data generator are allowed to take 5
values (\{$\pm2, \pm1,0\}$), and inference tries to fit data just
using ternary couplings ($\{\pm1,0\}$).

In the second series of simulations the patterns $\vec x\,^\mu$
are labeled according to their similarity to two vectors,  $\vec x\,_+$ and $\vec x\,_-$,
chosen as representative of the two classes.
In order to take into account the sparsity of the relevant genes,
we consider the variables $c_i=1$ if $ i=1,\dots k$ and $c_i=0$ if $ i=k+1,\dots N$, with $k \ll N$.
We thus classify the patterns according to the restricted Euclidean distance:
\begin{equation}
  y^\mu = \underset{z}{\operatorname{argmin}}\left( \sum_{i=1}^{N}c_i
  \sqrt{(  x_i^\mu -  x_{i,z} )^2} \right) = \underset{z}{\operatorname{argmin}}
\left(\sum_{i=1}^{k} \sqrt{(  x_i^\mu -  x_{i,z} )^2}\right) \;\;\;\;\; z\in \{+,-\} \ .
\end{equation}

\subsection{Results.}

We investigate the goodness of the two classifier, in the two
different situations described above, according to two different
measures. First, we consider the \textit{generalization error}. In
this case patterns are divided into disjoint training and test
sets. Learning is done on the training set, and the inferred
input-output rule is tested against the test set. The generalization
error is defined as the fraction of misclassified patterns in the test
set. Results are shown in Fig.~\ref{generr_art}.  For the first type
of dataset (perceptron-like) one can appreciate how BP outperforms the
simple centroids algorithm.  In the second type of dataset
(centroids-like), the centroids algorithm outperforms BP in the high
signature cases but the two algorithms are comparable for low
signatures.

As a second test on the accuracy of BP, we use the fact that the
data-generating signature is known. It can be compared directly to the
inferred one, allowing to group genes into four classes: True
positives (TP) are those genes which are contained in both signatures,
{\it i.e.} genes which are correctly identified as being relevant by
the inference algorithms. False positives (FP) are in the inferred
signature, but they are not in the true one. In analogy we define true
negatives (TN) as those which are in neither signature, and false
negatives (FN) are those genes which are in the data generator, but
they are not recognized by the algorithm. The {\it recall RC} (or
sensitivity) and the {\it precision PR} (or specificity) are thus
defined as:
\begin{equation}
RC=\frac{N_{TP}}{N_{TP}+N_{FN}} \ ,\ \ \ \ \ \  PR=\frac{N_{TP}}{N_{TP}+N_{FP}}\ .
\end{equation}
The recall $RC$ thus measure the fraction of correctly inferred genes
in the signature, while precision $PR$ is the fraction of inferred
couplings which are actually true ones. These two quantities are
obviously competing: An algorithm that includes all genes into the
signature has very low precision ($PR=\frac{N_{TP}}{N}$) but maximum
recall ($RC=1$), while including only the genes with very strong
signal into the signature may result in a good precision, but at the
cost of a potentially poor recall.  A perfect algorithm would have
recall and precision both equal to one, so the interplay between both
quantities is the relevant "observable". A curve
precision-versus-recall can be constructed by ranking all genes
according to their probability of having a non zero weight, and
introducing different cutoffs in the ranking: In Fig. \ref{roc_art} we
show the numerical results for the two datasets,
while a more detailed theoretical analysis
for the perceptron-like dataset can be found in \cite{key-martin}.
We see that for the data generated by the perceptron like rule, BP actually
performs considerably better than ranking by correlations, whereas results
are comparably good in the case of the centroid-like generator.

\section{Test on tumor data}
\label{sec:realdata}

As stated in the introduction, a common problem of micro-array
experiments is the small number of samples in the data set (from tens
to hundreds) compared with the number (thousands) of monitored genes.
Values of the parameter $\alpha=M/N$ in the considered data sets range
from $\alpha=0.01$ to $\alpha=0.01$. In principle these values of
$\alpha$ are below the threshold at which BP outperforms simple
pairwise correlation based methods, at least on data artificially
generated as described in the previous section and more extensively
discussed in \cite{key-martin}.  On the other hand, the simple ratio
between the number of patterns and the number of genes might not be
the relevant parameter on real data sets, due to the non trivial
correlations between different patterns, and between genes.
We first consider three data sets (leukemia, colon and prostate
cancer), already analyzed in \cite{Dettling} as benchmarks for other
algorithmic strategies ({\em e.g.~BagBoost, Random Forest, Support
Vector Machine, k-nearest neighbors, Diagonal Linear Discriminant
Analysis}). We further study a newer and larger data set on breast
cancer from two different laboratories \cite{van-de-vijver,
vant-veer}.

\begin{table}[htb]
\begin{tabular}{|p{2 cm}||*{6}{c|}}
\hline
& $N$ & $M$ & Class A/B & $M_\mathrm{training}$ & $M_\mathrm{test}$ & NDP\\
\hline
&&&&&&\\[-1.3em]
\hline
Leukemia & 3571 & 72 &  25/47 &48 & 24 & 200\\
\hline
Prostate & 6033 & 102 & 52/50 & 68  & 34 & 100\\
\hline
Colon  & 2000 &  62  &  40/22 &42 & 20 & 200\\
\hline
Breast & 6401 &  174 & 86/88  & 20-160 & 154-14 & 100\\
\hline
\end{tabular}
\protect\caption{In this table we display the number of probes $N$,
  the number of patterns $M$, the class composition of the data set
  Class A/B, the size of the training set $M_\mathrm{training}$, the
  size of the test set $M_\mathrm{test}$, and the number of different
  partitions NDP (see text) for each of the data set analyzed. }
\label{tab:dataset}
\end{table}

\subsection{Description of the data sets}

We have analyzed four different data sets of cancer tissues:

\begin{itemize}
\item {\em Leukemia}: It consists of 72 samples of two
  subtypes of leukemia -- 25 samples of acute myeloid leukemia (Class
  A) and 47 samples of acute lymphoblastic leukemia (Class B) --
  measured over 3571 genetic probes \cite{Golub}.

\item{\em Prostate} It consists of 102 samples -- 52 from prostate
  tumor tissues (Class A) and 50 from normal prostate tissues (Class
  B) -- measured over 6033 genetic probes \cite{Singh}.

\item {\em Colon}: It consists of 62 samples -- 40 from colon
  adenocarcinoma tissues (Class A) and 22 from normal colon tissues
  (Class B) -- measured over 2000 genetic probes \cite{Alon}.

\item{\em Breast}: The data set we have analyzed is the union of two
  different experiments presented in \cite{van-de-vijver, vant-veer}:
  it contains 311 samples measured over 24496 probes. We labeled
  the samples according to the following criterion: a metastasis event
  occurred in the first 5 years after the appearance of the tumor
  (Class P [poor prognosis]), the remaining samples did not develop a metastasis 
  in this time window and were labeled as Class G (good prognosis). In order
  to reduce the noise we removed all probes with nearly constant
  expression values across the data set. More specifically a probe is
  included in the data set if at least one of the of the following two
  conditions is met: (i) its variance is larger than 0.1, (ii) in at
  least ten samples its expression value is outside the window
  (-0.3,0.3). Eventually we ended up with 6401 probes.
  We further notice that the number of elements in Class P (86)
  is much smaller than those in Class G (225). Since a major aim in this
  context is to correctly classify all members of Class P (those developing 
  metastasis) and not to misclassify them as Class G cases, a larger influence 
  of class P data is needed. To obtain a more balanced dataset, we therefore 
  randomly removed 137 elements from Class G and we end up with a set of 174 
  patients.
\end{itemize}

In Tab.~\ref{tab:dataset} we resume the details of the data sets.

\subsection{Results}
\begin{figure}[htb]
\centerline{\includegraphics[width=8cm]{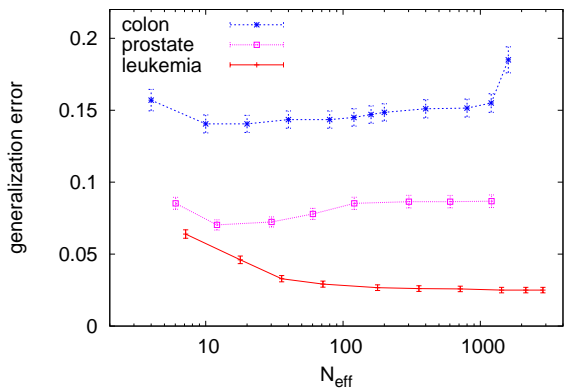}}
\caption{Generalization error as a function of $N_\mathrm{eff}$ in
the
  colon, prostate, and leukemia data set.}
\label{fig:generr_all}
\end{figure}

For each data set we construct different realizations of the training
and test sets by randomly permuting the $M$ samples. The first
$M_\mathrm{training}$ patterns will be the training set and the
remaining $M_\mathrm{test}=M-M_\mathrm{training}$ patterns will be
used as test set. In this way we are able to obtain results, which are
not dependent on a single specific arbitrary partitioning of the patterns 
into training and test sets, and to attribute statistical errors to measured
observables. In the last three columns of Tab.~\ref{tab:dataset} we
give the actual values of $M_\mathrm{training}$, $M_\mathrm{test}$,
and the number of different partitions for each of the data sets.

In all data sets discussed here, we have run BP starting at finite
temperature and using the cooling scheme described in Sec.~\ref{sec:mp},
which reached zero temperature. Running the algorithm at various fixed 
finite temperatures, we observed the generalization error to decrease
monotonously with temperature, and to saturate at low temperature. Here
we present directly the zero-temperature results obtained using cooling.

For the first three data sets (Leukemia, Prostate, and Colon) we first
run BP on the entire probe-set with the diluting field $h$ defined in
Eq.~(\ref{eq:gibbs}).  We show in Fig.~\ref{fig:generr_all} the
generalization error for the three data set as a function of
$N_\mathrm{eff}$, which, as already explained in Sec.~\ref{sec:mp},
can be fixed by tuning the field $h$.

In Fig.~\ref{fig:generr_all} two different scenarios emerge: the colon
and prostate curves display a minimum generalization error at a
relatively low value of $N_\mathrm{eff} \sim 10$, while the leukemia
curve is monotonously decreasing. This seems to indicate the presence
of a small set of probes relevant for the classification in the
prostate and colon case, while in the leukemia data set it seems that
all probes are relevant for the classification.

Let us recall that a given $N_\mathrm{eff}$ does not necessarily
indicate the actual size of the set of relevant probes. This would be
the case only when all probabilities $P_i(J)$ were completely
polarized ({\em i.e.}~$P_i(0)$ is either 0 or 1). Upon direct
inspection of our BP results it turns out that this is not the
case. No clear threshold is shown on the marginals $P_i(0)$.  The
dilution thus seems to be an effective strategy to attribute
differential weights to the probes, but it is not clear how to use it
to select a relevant signature in the data set.  A possible way is to
implement a decimation procedure as explained in Sec.~\ref{sec:mp}, in
order to test the performance of the algorithm on a restricted and
selected set of probes (signatures) of different size.

Domany {\em et~al.} in \cite{EinDor-Zuk-Domany2006} studied the
stability of different signatures as a function of the number of
samples used for learning.  They interestingly noted that the small
overlap between different signatures is not only due to the different
classification strategies used by different groups, but in particular
resulting from the small number of available cases. Such a
lack of robustness emerges even when using the same algorithm on data
generated with the same probabilistic framework
\cite{EinDor-Zuk-Domany2006}. We have investigated how stable are the
lists obtained with our decimation procedure.  It turns out that an
analogous phenomenon of instability occurs in our lists. We observed
very few genes appearing in all lists: the actual numbers differs
among the different data sets but it ranges from  10\% in the case of
Leukemia, to  0.02\% in the case of prostate tissues, where however the
number of relevant genes seems to be very small as we will see in the
following.


\begin{figure}
\includegraphics[width=0.32\columnwidth]{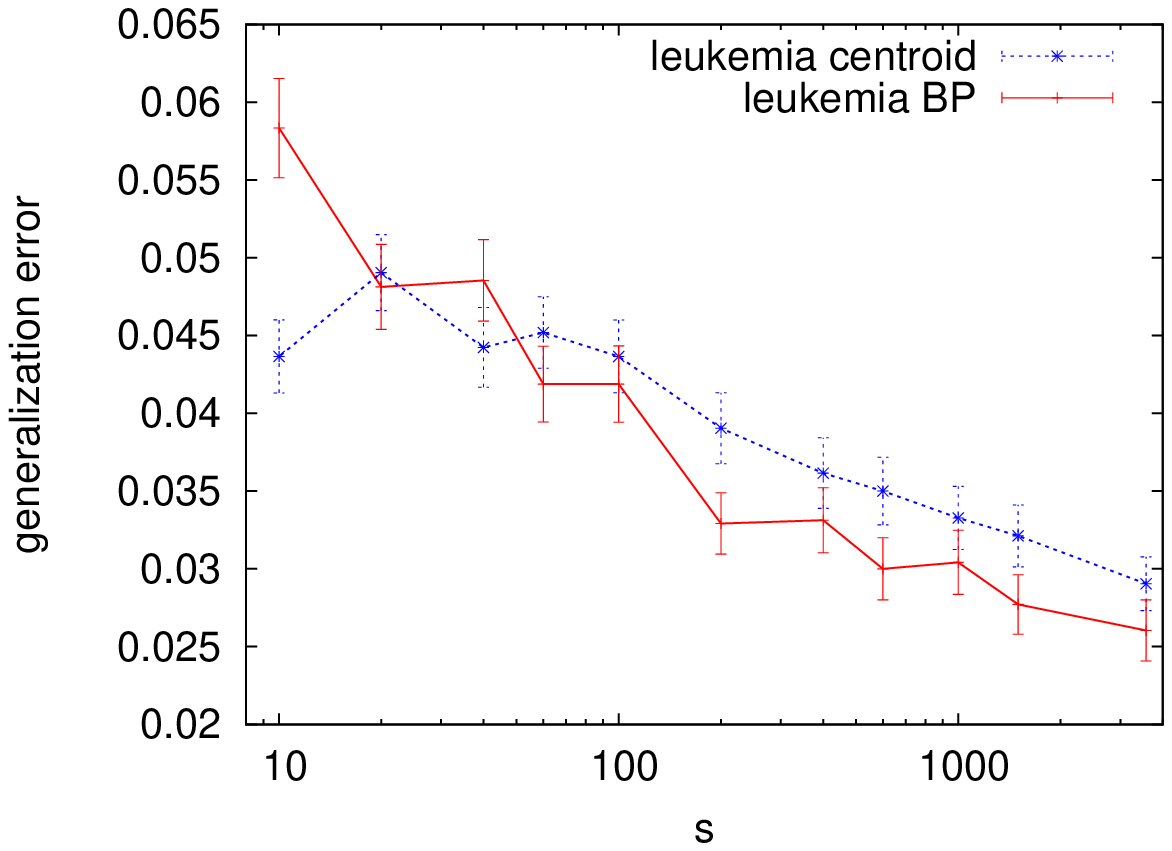}
\includegraphics[width=0.32\columnwidth]{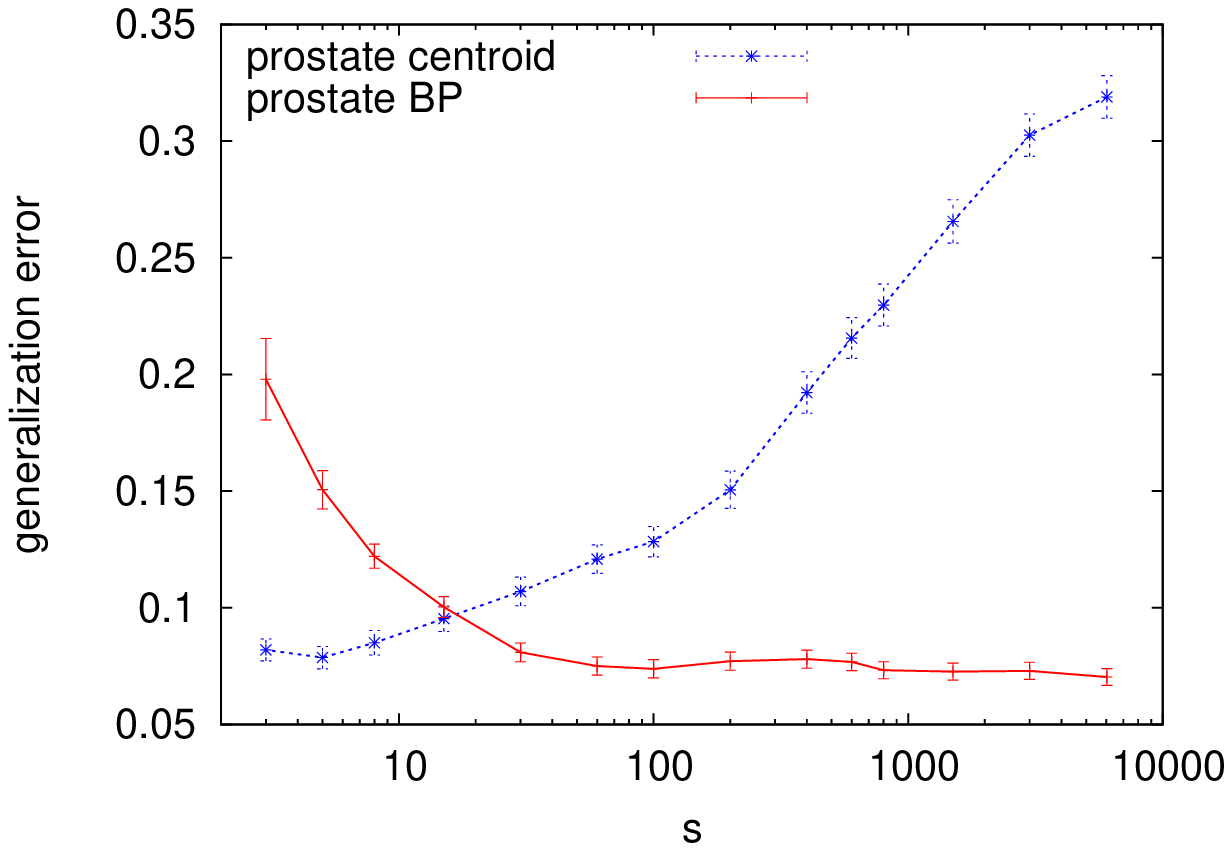}
\includegraphics[width=0.32\columnwidth]{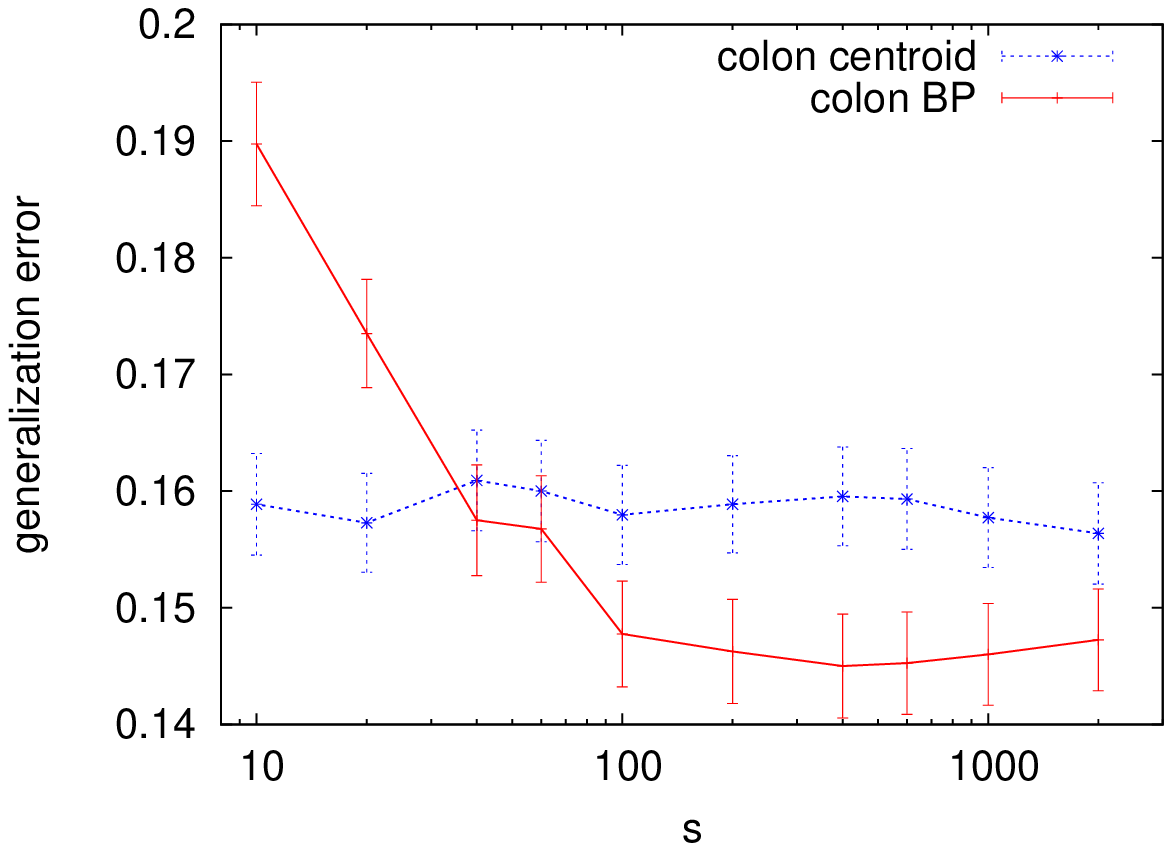}
\caption{Generalization error computed with Centroids (blue line) and
  Belief Propagation (red line), as a function of the signature size
  s, shown with a logarithmic scale on the x axes. We display results
  for leukemia (left panel), prostate (central panel), and colon
  (right panel).}
\label{fig:generr_3tumors}
\end{figure}

To understand how BP compares with simpler algorithmic strategies, we
used the centroids method selecting signatures of the same size. The
results for the first three data set are displayed in
Fig.~\ref{fig:generr_3tumors}. The curve for Leukemia displays a
monotonously decreasing profile in agreement with what we have
obtained fixing $N_\mathrm{eff}$ shown in
Fig.~\ref{fig:generr_all}. This seems to indicate that no signature
can be defined in this case.  The generalization error obtained with
the centroids algorithm is slightly worse then that of BP.  The
Prostate case displays dramatic difference in the two algorithms: from
the BP point of view the curve displays a plateau of minimal
generalization error for signature larger than 100, at odd with what
obtained fixing $N_\mathrm{eff}$ where a minimum is present around
$N_\mathrm{eff}=10$. Also in this case it seems difficult to determine
a clear signature in the data set. Centroids behaves in the
opposite way displaying a monotonously increasing function with
minimal generalization error at values of the signature lower than
10. The optimal generalization error achieved by the two algorithm in
this case is compatible.  The Colon case is analogous to that of
Prostate from the BP point of view (the minimum generalization error is obtained
for $N_\mathrm{eff}=10$ but for $s=400$), while centroids seems to be rather
insensitive to signature size in the whole interval analyzed. In this
case BP shows an overall better generalization ability.
The discrepancy between the values of the signature $s$ and of the $N_\mathrm{eff}$
for which the generalization error is minimum, comes from the
not unique set of possible relevant genes. While the algorithm is able
to select genes which are relevant and weight them adequately  ($N_\mathrm{eff}$ 
small), it is not able to define a single clear signature which alone is sufficient
for the classification. This implies  that, in the decimation procedure, a much 
higher number of genes, even if with small weight, is necessary to achieve a 
good predictive ability (so that the minimum generalization error is
reached for signatures $s$ much bigger than the probabilistic expectation
given by $N_\mathrm{eff}$). It is worth pointing out that the fact that BP 
behaves worse for very small sizes, is already present in the case
artificial data Fig.~\ref{generr_art}.

The best generalization errors of both BP and centroids are
displayed in Tab.~\ref{tab:generr} compared with the results
presented by Dettling in \cite{Dettling} where no statistical error
was associated with measures. We see that BP outperforms 4 out of 6
other algorithms on all three data sets, the other two algorithms perform
better on a single data set each.

\begin{table}
\begin{tabular}{|p{2 cm}||*{7}{c|}}
\hline
& {\bf BP} & centroids & BagBoost & RanFor & SVM & kNN & DLDA \\
\hline
&&&&&&&\\[-1.3em]
\hline
Leukemia & {\bf 0.025(2)} & 0.029(2) & 0.0408 & 0.025 & 0.035 & 0.0383 & 0.0292 \\
\hline
Prostate & {\bf 0.070(4)} & 0.071(4) & 0.0753  & 0.0788 & 0.0682 & 0.1059 & 0.1418  \\
\hline
Colon &   {\bf 0.140(6)}  &  0.156(4)  & 0.1610 &0.1543 & 0.1667 & 0.1638 & 0.1286  \\
\hline
\end{tabular}
\protect\caption{In this table we compare the generalization errors of our
  method (first column) against centroids, BagBoost, RanFor, SVM, kNN,
  and DLDA presented in \cite{Dettling}.}
\label{tab:generr}
\end{table}

\begin{figure}
\includegraphics[width=0.48\columnwidth]{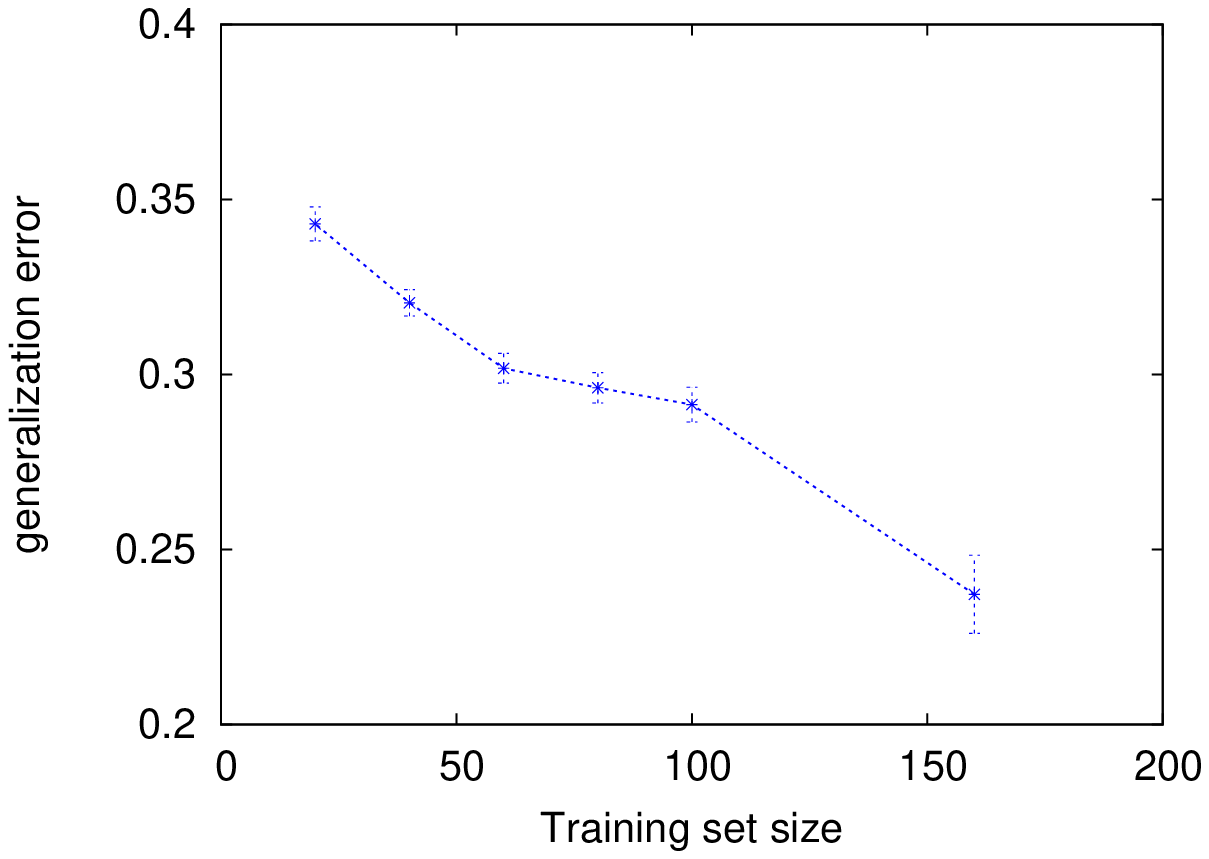}
\includegraphics[width=0.48\columnwidth]{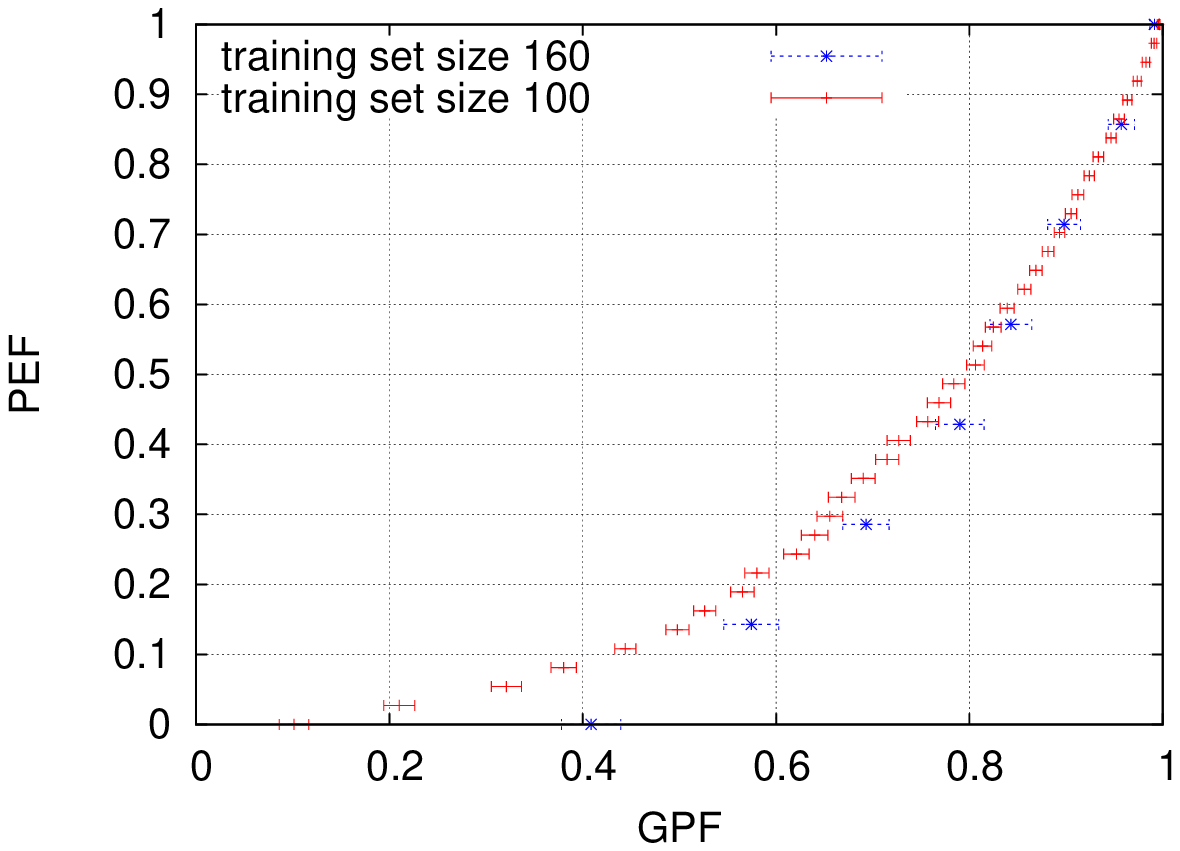}
\caption{Results for the breast-cancer dataset.
Generalization error as a function of the training set size (left)
and PEF (fraction of Class P patients classified as G) vs. 
GPF (fraction of correctly classified Class G patients).}
\label{fig:breast}
\end{figure}

In the breast-cancer case, we consider first the generalization error for various sizes 
of the training set ($M_{\rm training}=20,...,160$). The relatively large balanced data 
set containing 174 patients allows for the direct analysis of sample sizes ranging over
almost a full order of magnitude. The results are presented in the left panel of 
Fig.~\ref{fig:breast}, each point is averaged over 50 random selections of the training
set. We observe a strong monotonous decrease of the generalization error. This is a very 
encouraging sign and should motivate for collecting larger data sets. The generalization 
error of 30\% obtained for $M_{\rm training}=80$ is compatible with the one ($31\%$) 
reported in \cite{key-Michiels}.

However, in the breast cancer set we can go into more detail: The generalization error
treats good and poor cases (Classes G/P) in the same way. Indeed, one of the open 
challenges in breast cancer treatment is to recognize the correct cancer sub-type at
the earliest possible stage of disease, indicating, {\it e.g.}, the possible sensitivity
of the cancer to chemotherapy. In the present case, it is strongly preferable to 
erroneously include a Class G patient into Class P than vice versa; it has to be
avoided to predict the absence of metastasis for a patient who actually develops one, 
and possibly not to provide the necessary medical treatment to such a patient.

We therefore introduce four different subclasses:
\begin{enumerate}
\item patients in Class G (no metastasis) correctly classified as Class G, their number is called $M_{\rm GG}$;
\item patients in Class P (metastasis) misclassified as Class G, their number is called $M_{\rm PG}$;
\item patients in Class G (no metastasis) misclassified as Class P, their number is called $M_{\rm GP}$;
\item patients in Class P (metastasis) correctly classified as Class P, their number is called $M_{\rm PP}$.
\end{enumerate}
As said before, our primary aim is to classify correctly Class P patients. The fraction 
$M_{\rm PG}/(M_{\rm PP}+M_{\rm PG})$ of misclassified Class P patients has to be kept as small as possible; 
we refer to it as the {\it poor error fraction} (PEF). Once this is kept small, we also would like to
recognize correctly as many Class G patients as possible. The corresponding fraction in the
total Class G sample is $M_{\rm GG}/(M_{\rm GG}+M_{\rm GP})$, we refer to it as the {\it good prediction fraction} 
(GPF).

The GPF vs. PEF curve for a perfect classifier would be constantly equal to zero for the full
GPF range, while a random classifier would produce a curve PEF=GPF. Here we want to characterize
the relation of these two competing quantities (a constant prediction as P would have PEF=GPF=0, 
whereas a constant prediction G would have PEF=GPF=1), keeping however in mind that we want to 
keep the PEF low. The full curve is obtained done by changing the threshold $\tau$ defined in
Eq.~(\ref{eq:model}) of the inferred classifier.

The result is displayed in the right panel of Fig.~\ref{fig:breast} for training sets of size
$M_{\rm training}=100,\, 160$. For each possible PEF value we have averaged the curve over 50 
random balanced partitionings of our data into training / test sets. When training with 100 samples, 
we have 37 points corresponding to the possible numbers of misclassified Class P patients (0 to 36).
When training with 160 samples, this number reduced to 8 points. Note that the finding that the 
curve for larger training sets is located right of the other curve is consistent with the 
observation done in the left panel of Fig.~\ref{fig:breast}: Larger training sets
lead to better results. It is particularly striking that the PEF starts to grow much later,
the GPF at zero PEF grows from about 10\% to about 40\%. Again we see that larger data sets 
should be collected for developing more precise prognostic tools.

In order to compare our result with that of \cite{van-de-vijver}, we measured the value of the GPF 
corresponding to an average PEF value around $0.07$. This can be obtained by averaging over cases 
with $M_{\rm PG}$ being at most one. In this case we obtained a PEF of 0.076(1) and a GPF of 0.50(2). 
These values are comparable to \cite{van-de-vijver} where a PEF=0.071 and GPF=0.53 are reported
for a set of 180 patients and a leave-one-out procedure ($M_{\rm training}=M-1$).

In conclusion, for the breast cancer dataset, we obtain results comparable with state of the 
art studies, and we find a not saturated increase of performance with growing training
set size.

\section{Conclusion and Perspectives}
\label{sec:concl}
In this work we introduced and analyzed a message passing algorithm 
for the problem of classification and applied it to genome-wide 
expression patterns of cancer tissues. The aim of the algorithm is twofold:
on one hand one would like to get a reliable classifier, on the other
hand one wants to base this result on a minimal set of discriminating
genes. This additional requirement of parsimony is based on the
biological intuition that, as long as the expression level is
concerned, only a relatively small subset of genes is actually
significant for the development of the disease. Furthermore the
possibility of identifying a meaningful (possibly small) signature of
the disease will help the classification in two ways: (i)
computationally, the dimensional reduction of the patterns makes the
classification less prone to over-fitting, (ii) experimentally,
specific targeted gene-chip scanning could improve early stage cancer
diagnostic ({\em e.g.} in the case of breast-cancer).

The performance of our algorithm is found to be slightly better, although
comparable with the one of other state-of-the-art classification
techniques. Using three different data sets, the BP performed better
than 4 out of 6 algorithms on all data sets, and better than the other
2 algorithms on two out of the three data sets. The obtained results are 
compatible with the following theoretical intuition: The sparse quality 
(experimental and biological noise) and quantity (few expression patterns) 
of available data does not allow to extract substantially more information 
than the one seen also by simple algorithms. From this point of view the 
situation will improve in the near future since gene-chip technologies are 
becoming more and more diffused, and the emergence of new array technologies
based on longer oligo-nucleotides should reduce the experimental noise
of the expression measurements. This ongoing technological revolution
calls for the development of sophisticated global classification
techniques which are able to unravel, {\it e.g.}, combinatorial
control.

Another line of development in tumor detection is the integration of
different sources of information. From this point of view SNPs (Single
Nucleotide Polymorphism) appear to be the most promising high-throughput
genome-wide technology.

\section{Acknowledgments}
We thank Alfredo Braunstein, Yoshiyuki Kabashima, Michele Leone, and
Riccardo Zecchina for very interesting discussion on the problem of
inference and methodological aspects of message passing algorithms. We
also thank Enzo Medico for pointing out the Breast Cancer data set,
and the centroids algorithm.

\bibliography{biblio}

\end{document}